\documentstyle[prd,aps,preprint]{revtex}
\begin{document}
\textheight 21.5cm
\textwidth 15.5cm


\newcommand{\ls}[1]
   {\dimen0=\fontdimen6\the\font
    \lineskip=#1\dimen0
    \advance\lineskip.5\fontdimen5\the\font
    \advance\lineskip-\dimen0
    \lineskiplimit=.9\lineskip
    \baselineskip=\lineskip
    \advance\baselineskip\dimen0
    \normallineskip\lineskip
    \normallineskiplimit\lineskiplimit
    \normalbaselineskip\baselineskip
    \ignorespaces
   }

\def\a{\alpha}
\def\b{\beta}
\def\c{\chi}
\def\d{\delta} 			\def\D{\Delta}
\def\e{\epsilon}
\def\f{\phi} 			\def\F{\Phi}
\def\vf{\varphi}
\def\g{\gamma} 			\def\G{\Gamma}
\def\h{\eta}
\def\i{\iota}
\def\j{\psi} 			\def\J{\Psi}
\def\k{\kappa}
\def\l{\lambda} 		\def\L{\Lambda}
\def\m{\mu}
\def\n{\nu}
\def\o{\omega} 			\def\O{\Omega}
\def\p{\pi} 			\def\P{\Pi}
\def\q{\theta} 			\def\Q{\Theta}
\def\r{\rho}
\def\s{\sigma} 			\def\S{\Sigma}
\def\t{\tau}
\def\u{\upsilon} 		\def\U{\Upsilon}
\def\x{\xi} 			\def\X{\Xi}
\def\z{\zeta}
\def\sl{l}
\def\sf{f}
\def\hcpt{HHChPTh}

\def\tmax{t_{\rm max}}
\def\ms2{m_{B^*}^{2}}
\def\fr{\frac}
\def\ba{\begin{array}}
\def\ea{\end{array}}
\def\bz{\begin{equation}}
\def\ez{\end{equation}}
\def\by{\begin{eqnarray}}
\def\ey{\end{eqnarray}}
\def\ma{\matrix}
\def\nn{\nonumber}
\def\cd{\cdot}
\def\bkll{$B\to K^{(*)} \ell^+\ell^-$ }
\def\bqll{$b\to s \ell^+\ell^-$ }
\def\bks{$B\to K^* \ell^+\ell^-$ }
\def\bk{$B\to K \ell^+\ell^-$ }
\def\mll{m_{\ell\ell}}
\def\gev{{\rm GeV}}
\def\tl{\q_\ell}
\newtoks\slashfraction
	\slashfraction={.13}
\def\slash#1{\setbox0\hbox{$\, #1$}
	\setbox0\hbox to \the\slashfraction\wd0{\hss \box0}/\box0}
\def\nopagenumbers{\footline={\hfill}}

\tightenlines
\draft

\preprint{\vbox{\hbox{FERMILAB-Pub-95/113-T}
                \hbox{May 1995}}}
\vskip 0.5truecm

\title{Testing the Standard Model in $B\to K^{(*)} \ell^+\ell^-$  }
\author{Gustavo Burdman}
\vskip 0.5truecm
\address{Fermi National Accelerator Laboratory, P. O. Box 500, \\
Batavia, IL 60510, U.S.A. }

\maketitle
\ls{1.3}
\begin{abstract}
We study the potential of  $B\to K^{(*)} \ell^+\ell^-$ decays  as tests of the
standard model. After discussing the reliability of theoretical predictions for
the hadronic matrix elements involved, we examine the impact of different new
physics scenarios on various observables. We show that the angular information
in \bks together with the dilepton mass distribution can highly constrain new
physics. This is particularly true in the large dilepton mass region, where
reliable predictions for the hadronic matrix elements can be made with
presently available data. We compare the Standard Model predictions with those
of a Two-Higgs doublet model as well as TopColor models, all of which give
distinct signals in this region.
\end{abstract}


\vskip 0.5truecm
\newpage
\section{Introduction}
\label{sec:1}
Rare decays of $B$ mesons have a great potential as tests of the Standard Model
(SM). Processes involving Flavor Changing Neutral currents (FCNC) are of
particular interest given that in the SM they can only proceed via one or more
loops. This leads to very small rates allowing at the same time for possible
contributions from high energy scales to produce observable effects. For
instance, in the SM the top quark gives a very important, and actually
dominant, contribution to the process $b\to s\g$. Physics beyond the SM, e.g.
involving new heavy particles in the loop, could contribute with comparable
effects. In fact the observation by CLEO \cite{cleo} of the inclusive $b\to
s\g$ transition already constrains the parameter space of many theories
\cite{bsg_rev}. On the other hand, the process $b\to s\ell^+\ell^-$ involves
additional information on FCNC and is an important complement to $b\to s\g$ in
testing the SM. Although there is still no observation of any of these modes,
the experimental situation looks very promising both at $e^+e^-$ \cite{cleo_2}
as well as hadronic machines \cite{cdf}. Unlike in $b\to s\g$, where there are
no reliable ways to calculate the hadronic matrix elements in exclusive modes
(e.g. $B\to K^*\g$), the corresponding quantities in \bkll can be safely
predicted with the help of heavy quark symmetry \cite{iw_1}.
The fact that these exclusive modes can be theoretically clean together with
the additional information resulting from a richer kinematics, make these
decays very interesting both theoretically as well as experimentally. While
$b\to s\g$ proceeds via only one operator, the $b\to s\ell^+\ell^-$ processes
receive contributions from two additional operators at the weak scale. This
will imply that the information on higher energy scales,  encoded in the
coefficients of these operators, not only affects the rates but also the shape
of the dilepton mass and angular distributions.
These features have been recently pointed out by several authors
\cite{ali,ref_exc,liu}.
In this paper we clarify some the issues related to the extraction of hadronic
matrix elements in \bkll and also show explicitly how different signals probe
distinct aspects of the theories beyond the SM. In particular we show the
effects of a class of models in which the new physics couples preferentially to
the third generation \cite{chill}. In these type of scenarios, the large mass
of the top quark is explained by new gauge interactions that are strong with
the third generation, but rather weak with the first two. As a consequence, the
first low energy test of these models is in $B$ decays, with the  \bkll modes
the most   sensitives.
 In Sec.~\ref{sd} we present the short distance structure of $b\to
s\ell^+\ell^-$. We emphasize the distinction between theories that respect the
SM operator basis and those that expand it. This will prove to be relevant when
studying  the phenomenology of \bkll decays. In Sec.~\ref{hqs} we review the
relations predicted by heavy quark symmetry that allow reliable predictions of
\bkll decays in terms of experimental information from other modes.
In Sect.~\ref{predictions} we show the predictions of the SM, a Two-Higgs
doublet extension of it and of TopColor models, in order to compare the signals
of these different scenarios.
 We conclude in Sec.~\ref{conc}.
\section{Short Distance Structure}
\label{sd}
\begin{figure}
\vspace{8cm}
\includegraphics{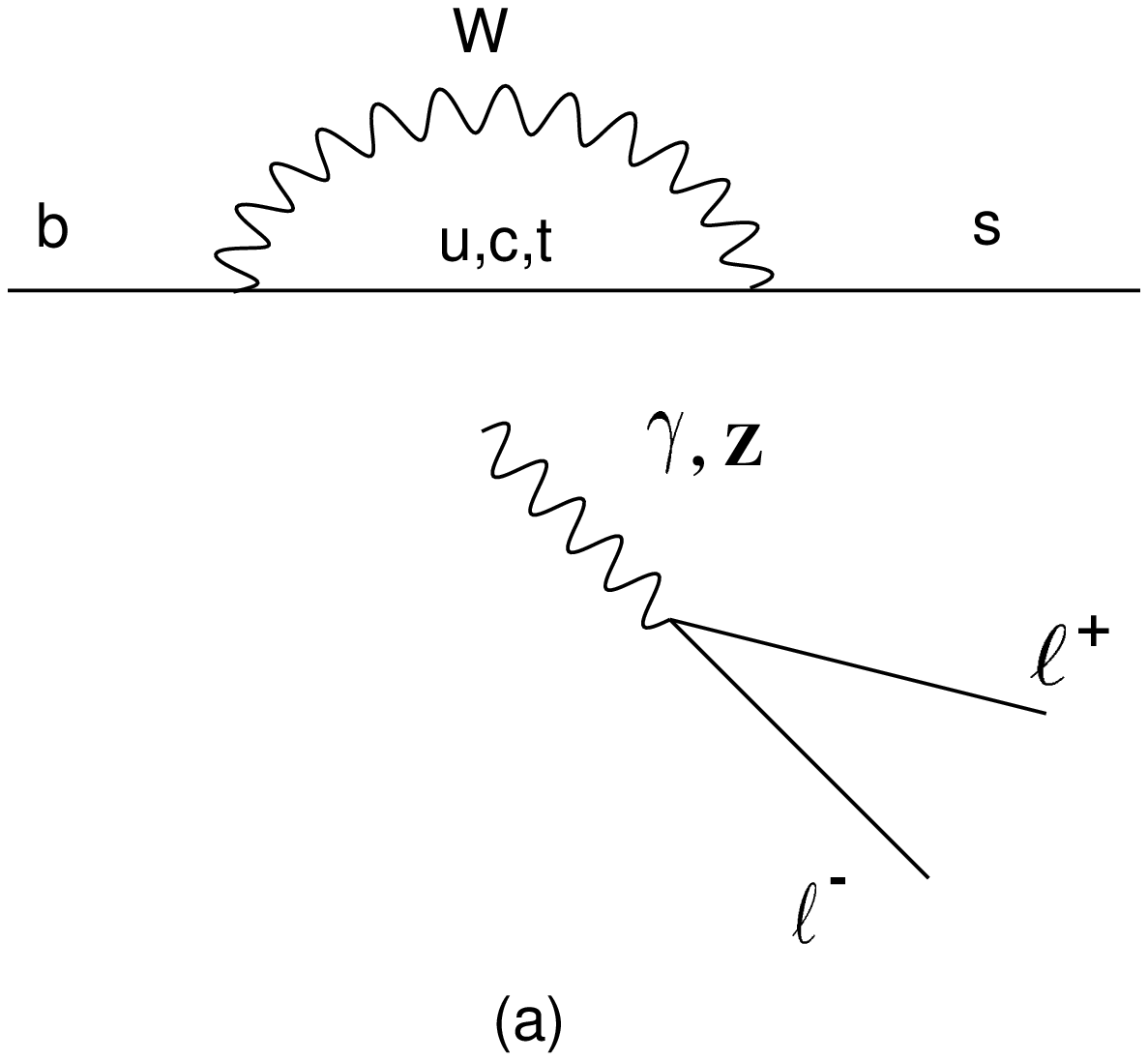}
\includegraphics{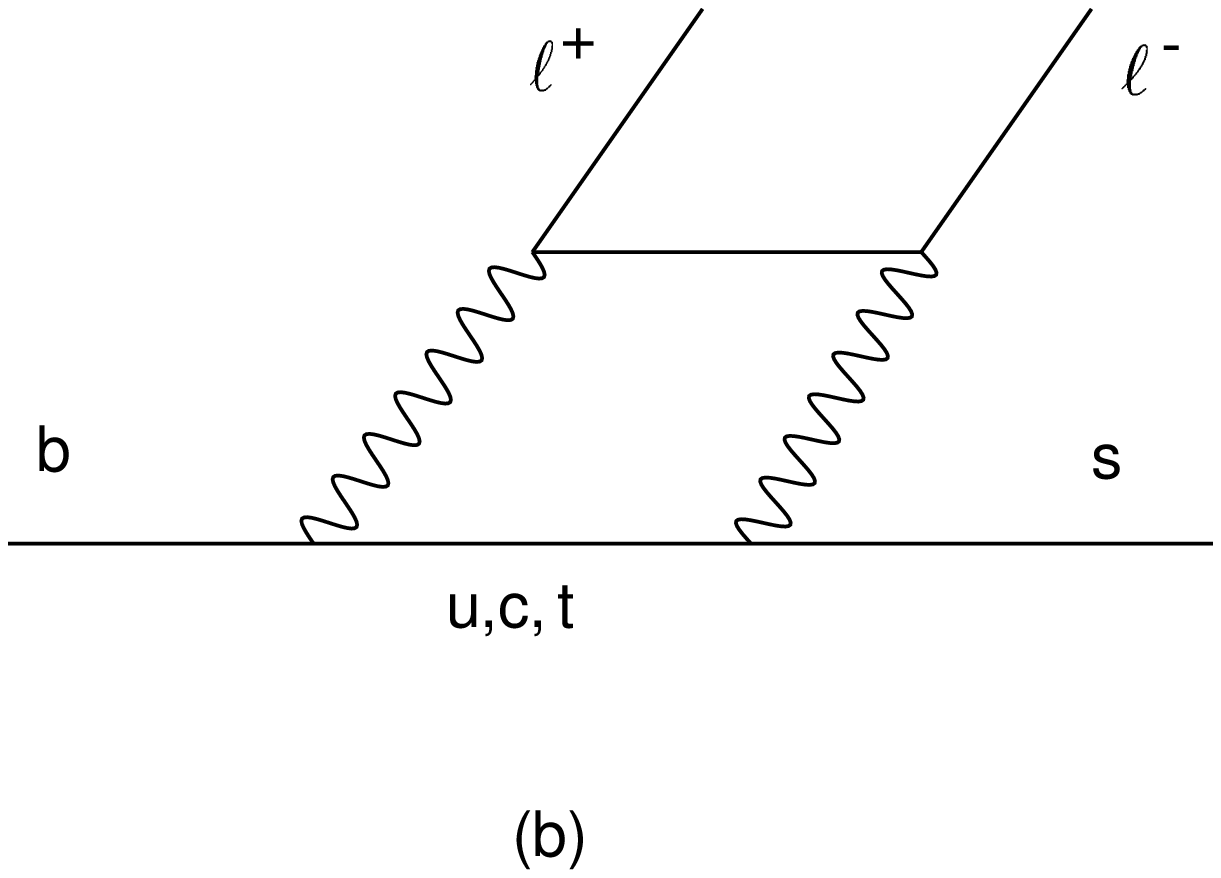}
\vspace{-1cm}\hspace{2.5cm}\hspace{8.0cm}\vspace{1cm}
\vspace{10cm}
\caption[]{Diagrams contributing to \bqll. }
\label{f_bsll}
\end{figure}

In the SM the diagrams contributing to \bqll processes are shown in
Fig.~\ref{f_bsll}. It is convenient to write down an effective theory at low
energies by integrating out the heavy degrees of freedom. These are the $W$
boson and top quark inside the loops in Fig.~\ref{f_bsll}. In general, in
theories beyond the SM there will be additional contributions. The effective
hamiltonian can be written as an operator product expansion as
\bz
H_{\rm eff}=\fr{4}{\sqrt{2}}G_F V_{tb}^{*}V_{ts}\sum_{i}C_i(\m)O_i(\m)
\label{heff}
\ez
where $\left\{O_i(\m)\right\}$ is the relevant operator basis and the $C_i(\m)$
are the corresponding coefficient functions which must cancel the dependence on
the energy scale $\m$. The dimension six operator basis is, in the SM
\cite{opbasis,grins_bsg}
\by
O_1&=&\left(\bar{s}_\a\g_\m b_{L\a}\right)\left(\bar{c}_\b\g^\m c_{L\b}\right)
\quad\quad\quad\quad\quad\;
O_6=\left(\bar{s}_\a\g_\m b_{L\b}\right)\sum_{q}\left(\bar{q}_\b\g^\m
q_{R\a}\right) \nn \\
O_2&=&\left(\bar{s}_\a\g_\m b_{L\b}\right)\left(\bar{c}_\b\g^\m c_{L\a}\right)
\quad\quad\quad\quad\quad\;
O_7=\fr{e^2}{16\p^2} m_b\left(\bar{s}\s_{\m\n}b_R\right) F^{\m\n}\nn\\
O_3&=&\left(\bar{s}_\a\g_\m b_{L\a}\right)\sum_{q}\left(\bar{q}_\b\g^\m
q_{L\b}\right)
\quad\quad\quad\quad
O_8=\fr{e^2}{16\p^2} \left(\bar{s}\g_\m
b_L\right)\left(\bar{\ell}\g^\m\ell\right)\label{opbasis}\\
O_4&=&\left(\bar{s}_\a\g_\m b_{L\b}\right)\sum_{q}\left(\bar{q}_\b\g^\m
q_{L\a}\right)
\quad\quad\quad\quad
O_9=\fr{e^2}{16\p^2}\left(\bar{s}\g_\m
b_L\right)\left(\bar{\ell}\g^\m\g_5\ell\right) \nn\\
O_5&=&\left(\bar{s}_\a\g_\m b_{L\a}\right)\sum_{q}\left(\bar{q}_\b\g^\m
q_{R\b}\right) \nn
\ey
\noindent where $q_{L,R}=\fr{(1\mp\g_5)}{2}q$ and $\a, \b$ are color indices.
The coefficients at $\m=m_b$ are obtained when  their values at $\m=M_W$ are
evolved using the renormalization group equation. This has been done in  a
partial leading logarithmic approximation in \cite{opbasis} for the SM as well
as for Two-Higgs doublet models. The complete leading logarithmic approximation
for the SM has been recently computed in \cite{com_ll}. For the purpose of this
paper it is sufficient to make use of the approximation in \cite{opbasis},
given that the uncertainties related to long distance dynamics, even when
reduced by symmetry arguments (see next section), are still larger than the
error made by using it. For a complete discussion see \cite{bur_ee}.
 The short distance information is then encoded in the $C_i(M_W)$'s. Several
theories beyond the SM share this operator basis with it. Their extra
contributions at the electroweak or higher energy scales appear as changes in
the values of the coefficients at $\m=M_W$.
There are also theories that require an extension of the operator basis. For
instance, the chirality of the quark currents could be reversed, giving
\begin{mathletters}
\label{w_basis}
\bz
O'_7=\fr{e^2}{16\p^2} m_b\left(\bar{s}\s_{\m\n}b_L\right)
F^{\m\n}\label{w_basis_1}
\ez
\bz
O'_8=\fr{e^2}{16\p^2} \left(\bar{s}\g_\m
b_R\right)\left(\bar{\ell}\g^\m\ell\right) \label{w_basis_2}
\ez
\bz
O'_9=\fr{e^2}{16\p^2}\left(\bar{s}\g_\m
b_R\right)\left(\bar{\ell}\g^\m\g_5\ell\right) \label{w_basis_3}
\ez
\end{mathletters}
as well as the ones resulting from $b_L\to b_R$ in $O_1-O_6$.

\noindent The SM, supersymmetry, multi-Higgs models and most technicolor
scenarios only modify the value of the short distance coefficients $C_i(M_W)$,
without inducing additional operators. On the other hand, left-right symmetric
models, compositness and topcolor are among the theories capable of inducing
sizeable wrong-chirality contributions as well as shifts in the
normal-chirality coefficients. These two very different groups of electroweak
symmetry breaking scenarios are likely to give drastically different signals,
for instance for the angular distributions. The only operator entering in $b\to
s\g$ at the weak scale is $O_7$. Therefore, radiative processes have a rather
limited power as SM tests: only the shifts induced in $C_7(M_W)$ are probed.
Conversely, \bqll processes are more sensitive to new physics which might give
small or null
deviations of the $b\to s\g$ rate but still give large contributions to the
other coefficients in (\ref{heff}). These would produce not only a different
rate but also distinct patterns in the dilepton mass distributions as well as
the angular distributions.

\section{Reliable Predictions for \bkll}
\label{hqs}
The use of experimental information in \bqll processes requires the theoretical
understanding of hadronic effects. The inclusive rate $B\to X_s\ell^+\ell^-$ is
theoretically clean in this respect and can be predicted with rather low
uncertainties \cite{ali,falk}. However, its experimental observation might
prove to be a very difficult task. On the other hand, exclusive modes like
$B\to K\ell^+\ell^-$ and $B\to K^*\ell^+\ell^-$ are more accesible to both
$e^+e^-$ and hadronic machines, but the need to compute hadronic matrix
elements of the operators in (\ref{opbasis}) and (\ref{w_basis}) introduces
potentially large hadronic uncertainties. The situation appears to be similar
to the one in $B\to K^*\g$, where the large disagreement among model
calculations of the matrix element of $O_7$ between $B$ and $K^*$ renders this
mode almost useless as a test of the SM. However, in the present case these
large uncertainties can be avoided by making use of Heavy Quark Symmetry (HQS),
which relates the hadronic matrix elements entering in \bkll decays to those
entering in semileptonic decays \cite{iw_1}. It must be emphasized that we {\em
do not} consider the limit in which the $s$ quark is heavy, as is done in part
of previous work on these decays.
In the rest of this section we review the implementation of these relations and
make use of them in \bqll exclusive decays.

\subsection{$B\to K^*\ell^+\ell^-$}
As we will see below, this mode is the most interesting one from the
phenomenological point of view. We need the matrix elements of the operators
$O_7$, $O_8$, $O_9$, $O'_7$, $O'_8$ and $O'_9$. The hadronic matrix elements of
$O_7$ and $O'_7$ can be written as
\by
\langle K^*(k)|\bar{s}\s_{\m\n}(1\pm\g_5) b|B(P)\rangle &=&
\e_{\m\n\a\b}\left\{A \e^{*\a}P^\b +B\e^{*\a}k^\b + C\e^*.P P^\a k^\b\right\}
\nn \\
& &\pm i\left\{A\left(\e^{*\m}P^\n - \e^{*\n}P^\m\right) +B\left(\e^{*\m}k^\n -
\e^{*\n}k^\m\right)\right. \nn \\
& & + \left. Ce^*.P\left(P^\m k^\n-P\n k^\n\right)\right\} \label{bk_sig}
\ey
where $A$, $B$ and $C$ are unknown functions of $q^2=(P-k)^2$, the squared
momentum transferred to the leptons. On the other hand, the matrix elements of
the semileptonic operators $O_8$, $O_9$, $O'_8$ and $O'_9$ are parametrized by
\by
\langle K^*(k)|\bar{s}\g_\m (1\pm\g_5) b|B(P)\rangle &=&
i g \e_{\m\n\a\b} \e^{*\n}(P+k)^\a (P-k)^\b \nn \\
& &\pm f \e^{*}_{\m}\pm a_+ e^*.P (P+k)_\m\pm a_- e^*.P (P-k)_\m \label{bk_sem}
\ey
with $g$, $f$ and $a_{\pm}$ also unknown functions of $q^2$.

\noindent In the Heavy Quark Limit, where the mass of the $b$ quark is
infinitely heavy compared to the typical scale of the strong interactions, the
Dirac structure of (\ref{bk_sig}) is related to that of (\ref{bk_sem}). The
simplification arises because in the $m_b \gg \L_{\rm QCD}$ limit the heavy
quark spinor loses its two lower components becoming
\bz
b(x)\approx \left(\ba{c} \Psi(x) \\ 0\ea\right) \label{spinor}
\ez
That is, the two lower components are suppressed by $m_b$. As a consequence,
the action of gamma matrices on the $b$ spinor simplifies. One has, for
instance $\g_0 b=b$. The $(0i)$ component of (\ref{bk_sig}) is now related to
the $(i)$ component of (\ref{bk_sem}) by the relations
\by
i\s_{0i}&=&\g_i \nn \\
& & \label{dirac}\\
i\s_{0i}\g_5&=&-\g_i\g_5 \nn
\ey
By making use of (\ref{dirac}) we can obtain now relations among the
form-factors in (\ref{bk_sig}) and (\ref{bk_sem}). They are \cite{iw_1}
\begin{mathletters}
\label{ffrel}
\bz
A=\fr{-f+2m_B k_0 g}{m_B} \label{rel_1}
\ez
\bz
B=-2 m_B g \label{rel_2}
\ez
\bz
C=\fr{a_+-a_- +2g}{m_B} \label{rel_3}
\ez
\end{mathletters}
Furthermore, the semileptonic $B\to K^*$ from-factors $g$, $f$ and $a_\pm$ are
identical to the $B\to\rho\ell\nu$ form-factors in the $SU(3)$ limit.
Therefore, the hadronic matrix elements entering in $B\to K^*\ell^+\ell^-$ are
given by the $B\to\rho\ell\nu$ from-factors to leading order in $\L_{\rm
QCD}/m_b$ and in the $SU(3)$ limit. In this way, experimental information on
$B\to\rho\ell\n$ can be used as a prediction for $B\to K^*\ell^+\ell^-$.
The relations (\ref{ffrel}) can be also used to predict the $B\to K^*\g$ rate.
In this case, however the region of the $B\to\rho\ell\nu$ Dalitz plot that can
be used is negligibly small. This is due the fact that one needs the
semileptonic form-factors evaluated at $q^2=0$ in the case with a real photon.
Moreover, the $\rho$ must be left-handed in order to have the same combination
of form-factors that appear in the radiative decay. Both conditions can only be
satisfied in one corner, corresponding to the  $\rho$ and the charged lepton
having both  maximum momenta. The rate at this point vanishes due to phase
space, and thus an extrapolation to this point must be made in order to extract
a model independent quantity that would constitute the $B\to K^*\g$ hadronic
matrix element \cite{bd_kg}.
The situation is very different here. We are able to use all the
$B\to\rho\ell\nu$ Dalitz plot, which in the SU(3) limit covers exactly the
$B\to K^*\ell^+\ell^-$ Dalitz plot.

\noindent Presently, there are only upper limits for $B\to\rho\ell\nu$
\cite{cleo_3}. However the observation of $B\to\p\ell\n$ at CLEO \cite{cleo_3}
indicates that there will be data on this mode very soon.  We can make use of
present data by taking an extra step in using the heavy quark symmetries. This
time we use the flavor symmetry that relates processes with $b$ and $c$ quark
hadrons. This implies relations among the semileptonic $B\to K^*$ form-factors
entering in
(\ref{bk_sem}) and those in the decay  $D\to K^* \ell\n$. Up to a
short-distance correction factor, these relations are the simple mass scaling
laws \cite{iw_1}:
\begin{mathletters}
\label{scaling}
\bz
f^{B}(v.k)=\sqrt{\fr{m_B}{m_D}}f^{D}(v.k) \label{scal_1}
\ez
\bz
g^{B}(v.k)=\sqrt{\fr{m_D}{m_B}}g^{D}(v.k) \label{scal_2}
\ez
\bz
(a_+-a_-)^{B}(v.k)=\sqrt{\fr{m_D}{m_B}}(a_+-a_-)^{D}(v.k) \label{scal_3}
\ez
\end{mathletters}
\noindent where the form-factors in the left-hand side of (\ref{scaling}) are
those in $B\to K^*\ell^+\ell^-$ and those in the right-hand side those entering
in $D\to K^*\ell\n$. The relations (\ref{scaling}) are valid when the
form-factors are evaluated at the same value of $v.k$, with $v$ the heavy meson
four-velocity and $k_\m$ the $K^*$ four-momentum. In the rest frame of the
heavy meson this is the $K^*$ recoil energy, $E_{K^*}$. Thus, we can only
predict \bks for values of $E_{K^*}$ ranging from $m_{K^*}$ to the maximum
recoil in $D\to K^*\ell\n$. In terms of $q^2=\mll^2$, the dilepton invariant
mass squared, this implies the  window $4.0\gev <\mll <4.4\gev$, where the
upper value is actually the maximum allowed dilepton mass. Model independent
predictions for the
\bks  form-factors for smaller dilepton masses will only be possible when the
$B\to\rho\ell\n$ data is available. We will make use of the form-factors
extracted in $D\to K^*\ell\n$ decays \cite{dklnu} to illustrate how the SM
predictions compare with those of some of its extensions.

In order to take full advantage of the information provided by this decay we
first notice that the amplitude can be divided into two non-interfering pieces
corresponding to left and right-handed leptons. Each of them can then be
expressed as a sum of helicity amplitudes. The squared  amplitude  is
\bz
\left| A\left(B\to K^* \ell ^+\ell ^- \right)\right|^2=
|H_{+}^{L}|^2+|H_{-}^{L}|^2 +|H_{+}^{R}|^2+|H_{-}^{R}|^2 +|H_{0}|^2\label{amp}
\ez
where the $L(R)$ refer to left (right)-handed leptons and the subindices $\pm$
and $0$ indicate the transverse and longitudinal polarizations of the $K^*$.
When the heavy quark relations (\ref{ffrel}) are imposed, the helicity
amplitudes depend only on the semileptonic form-factors $f$, $g$ and $a_+$. The
transverse helicities take the form
\begin{mathletters}
\label{helamps}
\by
H_{\a}^{L}&=& -\left[C_7\fr{m_b\left(m_B-E_*+\h_\a{\bf
k}\right)}{q^2}+\fr{C_8-C_9}{2}\right]\left(f+\h_\a 2m_B{\bf k}g\right) \nn \\
& &+\left[C'_7\fr{m_b\left(m_B-E_*-\h_\a{\bf
k}\right)}{q^2}+\fr{C'_8-C'_9}{2}\right]\left(f-\h_\a 2m_B{\bf k}g\right)
\label{helamp_1}
\ey
\by
H_{\a}^{R}&=& -\left[C_7\fr{m_b\left(m_B-E_*+\h_\a{\bf
k}\right)}{q^2}+\fr{C_8+C_9}{2}\right]\left(f+\h_\a 2m_B{\bf k}g\right) \nn \\
& &+\left[C'_7\fr{m_b\left(m_B-E_*-\h_\a{\bf
k}\right)}{q^2}+\fr{C'_8+C'_9}{2}\right]\left(f-\h_\a 2m_B{\bf k}g\right)
\label{helamp_2}
\ey
\end{mathletters}
\noindent where $\a=+,-$ and $\h_a=(1,-1)$. The longitudinal helicity is
\by
H_0&=&\fr{m_{B^2}}{2m_{K^*}\sqrt{q^2}}\left\{-2(C_7-C'_7)\fr{m_b}{q^2m_B}\left[f(m_BE_*-m_{K^*}^{2})+2m_{B}^{2}{\bf k}^2g\right] \right. \nn \\
& &+\left. \fr{(C_8-C_9-C'_8+C'_9)}{2} \left[4{\bf
k}^2a_+-\fr{2E_*}{m_B}f\right]\right\}\label{hel_long}
\ey
In (\ref{helamps}) and (\ref{hel_long}) $E_*$ and ${\bf k}$ are the energy and
spatial momentum of the $K^*$ in the $B$ rest frame, the short distance
coefficients must be evaluated at $\m=m_b$ and the form-factors at
$q^2=\mll^2$.
The latter can be either the semileptonic form-factors extracted from
$B\to\rho\ell\n$ or, by applying (\ref{scaling}), those from $D\to K^*\ell\n$.
One can immediately see from (\ref{helamps}) that there is an interesting
interplay between the short distance coefficients and the combination of
form-factors entering in each transverse helicity. For instance, the first term
in each of (\ref{helamp_1}) and (\ref{helamp_2}) corresponds to the
contributions from normal-helicity operators, whereas the second line comes
from the one with wrong chirality. Depending on the sign $\h_\a$, some
contributions will be enhanced and some suppressed by the form-factors. This
will lead to rather distinct signals when considering different physics
scenarios at $M_W$ or higher scales.

The angular information is most sensitive to the details of the helicity
amplitudes. The forward-backward asymmetry for leptons was considered in the
contexts of both inclusive \cite{ali} and exclusive \cite{liu} \bqll processes.
Defining $\tl$ as the polar angle formed by the $\ell^+$ and the $B$ meson in
the $\ell^+\ell^-$ rest frame, the double differential decay rate has the form
\by
\fr{d^2\G}{dq^2
d\cos\tl}&=&\fr{G_{F}^{2}\a^2\left|V^{*}_{tb}V_{ts}\right|^2}{768\p^5m_{B}^{2}}kq^2\left\{\left(1+\cos\tl\right)^2\left[|H_{+}^{L}|^2+|H_{-}^{R}|^2\right]\right. \nn \\
&
&\left.\left(1-\cos\tl\right)^2\left[|H_{-}^{L}|^2+|H_{+}^{R}|^2\right]+2\sin^2\tl|H_{0}|^2\right\} \label{dg2}
\ey
Then the forward-backward asymmetry is defined by
\bz
A_{FB}(q^2)=\fr{\int_{0}^{1}\fr{d^2\G}{dq^2 d\cos\tl}dq^2 -
\int_{-1}^{0}\fr{d^2\G}{dq^2 d\cos\tl}dq^2}{\int_{0}^{1}\fr{d^2\G}{dq^2
d\cos\tl}dq^2 + \int_{-1}^{0}\fr{d^2\G}{dq^2 d\cos\tl}dq^2} \label{afb}
\ez
which, making use of (\ref{dg2}), gives
\bz
A_{FB}=\fr{3}{4}\quad\fr{|H_{-}^{L}|^2+|H_{+}^{R}|^2-|H_{+}^{L}|^2-|H_{-}^{R}|^2}{|H_{-}^{L}|^2+|H_{+}^{R}|^2+|H_{+}^{L}|^2+|H_{+}^{R}|^2+|H_{0}|^2}
\label{afb_hel}
\ez
Alternatively, one can simply use the angular distribution. In the next section
we present results for the asymmetry and the dilepton mass distributions in
several theories, including the SM.

\subsection{\bk}
 We now turn to the pseudosclar decay mode.
The hadronic matrix elements of the operators $O_i$ and $O'_i$, $i=7,8,9$, are
now
\bz
\langle K(k)|\bar{s}\s_{\m\n} b|B(P)\rangle = D\left[(P+k)_\m (P-k)_\n -
(P+k)_\n (P-k)_\m\right] \label{bp_sig}
\ez
\bz
\langle K(k)|\bar{s}\g_\m b|B(P)\rangle = f_+ (P+k)_\m +f_-(P-k)_\m
\label{bp_sem}
\ez
with $D$, and $f_{\pm}$ unknown functions of $q^2=\mll^2$. Using (\ref{dirac})
one obtains
\bz
D=-\fr{(f_+-f_-)}{2 m_B} \label{ffrel_p}
\ez
In the SU(3) limit $f_{\pm}$ are the $B\to\p\ell\n$ form-factors. Measurements
of this mode already exist \cite{cleo_3} but are still not precise enough. On
the other hand the flavor heavy quark symmetry predicts \cite{iw_1}
\bz
(f_+-f_-)^{B}(v.k)=\sqrt{\fr{m_B}{m_D}} (f_+-f_-)^{D}(v.k) \label{scal_p}
\ez
where the $B$ and the $D$ labels refer to the \bk and $D\to K\ell\n$ decays
respectively. The window in which a reliable prediction for \bk can be obtained
from the charm semileptonic decay data is now $4.2 \gev < \mll < 4.8 \gev$.
The dilepton mass distribution takes the form
\bz
\fr{d\G}{dq^2}=\fr{G_{F}^{2}|V_{tb}^{*}V_{ts}|^2\a^2{\bf
k}^2}{192\p^5}\left\{\left|(C_8+C'_8)f_+-2Dm_b(C_7+C'_7)\right|^2+\left|(C_9+C'_9)f_+\right|^2\right\}
\label{dq2k}
\ez
{}From (\ref{dq2k}) it can be seen that the chirality of the operators is not
tested in this mode. It is not possible to distinguish a shift in the
coefficients of the normal operator basis from a non-zero value of the ``wrong"
chirality operators in (\ref{w_basis}). The information of this decay could be,
however, an important complement to \bks.

\section{Predictions}
\label{predictions}
In this section we consider various Electroweak Symmetry Breaking Scenarios and
their signals in \bkll decays. The main purpose is to illustrate the potential
of these rare decays to discriminate among theories. The results presented here
make use of the form-factor relations (\ref{ffrel}) {\em and} (\ref{scaling}).
The use of the latter implies a restriction of the theoretically safe
predictions to the region with $\mll >4.0\gev$ in the \bks case. We slightly
extend this region down to $3.8$ GeV to cover the dilepton mass region above
the $\Psi'$ background from the indirect process $B\to K^*\Psi'\to
K^*\ell+\ell^-$. For the  \bk case we present results for the region $4.2 {\rm
GeV} <\mll<4.8 \gev$, which is allowed by the use of $D\to K\ell\n$ data in
(\ref{scal_p}).

\subsection{Standard Model}
At the electroweak scale the only operators contributing to the \bqll
transitions are $O_7$, $O_8$ and $O_9$. However, when evolved down to the $b$
quark scale, they mix with $O_1$ and $O_2$. The coefficients at $\m=M_W$ are
\cite{coeff_mw}
\by
C_7(M_W)&=&\fr{1}{2}A(x) \label{c7mw} \\
C_8(M_W)&=&\fr{B(x)}{s^2\q_w}+\fr{4s^2\q_w-1}{s^2\q_w}C(x)+D(x)-\fr{4}{9}
\label{c8mw}\\
C_9&=&-\fr{B(x)}{s^2\q_w}+\fr{C(x)}{s^2\q_w} \label{c9sm}
\ey
where $x=(m_{t}^{2}/M_{W}^{2})$ and with
\by
A(x)&=&\fr{x}{(x^3-1)}\left\{\fr{2}{3}x^2+\fr{5}{12}x-\fr{7}{12}-\fr{x\left(3/2x-1\right)}{(x-1)}\ln x\right\} \label{ax}\\
B(x)&=&\fr{x}{4(x-1)}\left\{\fr{1}{(x-1)}\ln x -1\right\} \label{bx}\\
C(x)&=&\fr{x}{4(x-1)}\left\{\fr{x}{2}-3+\fr{(3/2x^2+1)}{(x-1)}\ln x\right\}
\label{cx}\\
D(x)&=&\fr{1}{(x^3-1)}\left\{\fr{25}{36}x^2-\fr{19}{36}x^3+\fr{\left(-\fr{x^4}{6}+\fr{5}{3}x^3-3x^2+\fr{16}{9}x-\fr{4}{9}\right)}{(x-1)}\ln x \right\}\label{dx}
\ey
We also need $C_1(M_W)=0$ and $C_2(M_W)=-1$.
Defining $\h=\a_s(m_b)/\a_s(M_W)$, the evolution to $m_b$ gives \cite{opbasis}
\bz
C_7(m_b)=\h^{-16/23}\left\{C_7(M_W)-\fr{58}{135}\left(\h^{10/23}-1\right)C_2(M_W)-\fr{29}{189}\left(\h^{28/23}-1\right)C_2(M_W)\right\}, \label{c7mb}
\ez
\by
C_8(m_b)&=&C_8(M_W)+\fr{4\pi}{\a_s(M_W)}\left\{\fr{4}{33}\left(\h^{-11/23}-1\right) \right.\nn\\
& &\left.+\fr{8}{27}\left(1-\h^{-29/23}\right)\right\}C_2(M_W)
+\left[C_1(m_b)+C_2(m_b)\right]g(m_c/m_b;q^2) \label{c8mb}
\ey
where
\bz
C_{1,2}(m_b)=\fr{1}{2}\left[\h^{-6/23}\mp\h^{12/23}\right]C_2(M_W)
\label{c12mb}
\ez
and the last term in $C_8(m_b)$ comes from the one-loop contributions of $O_1$
and $O_2$, which gives a dependence on $q^2$. For $q^2>4m_{c}^{2}$ it is given
by the function
\bz
g(z,s)=-\left\{\fr{4}{9}\ln z^2-\fr{8}{27}-\fr{16}{9}\fr{z^2}{s}
+\fr{2}{9}\sqrt{1-\fr{4z^2}{s^2}}\left(2+\fr{4z^2}{s^2}\right)\left[\ln\left|\fr{1+\sqrt{1-\fr{4z^2}{s^2}}}{1-\sqrt{1-\fr{4z^2}{s^2}}}\right|+i\p\right]\right\} \label{gzs}
\ez
where the imaginary part arises when the charm quarks in the loop go on shell.
In calculating the asymmetry and dilepton mass distribution we do not include
the long distance pieces coming from the $B\to  K^*J/\Psi$ plus
$J/\Psi\to\ell^+\ell^-$ or the analogous process for the $\Psi'$. The
limitations from using $D\to K^*\ell\n$ data keep our predictions at $\mll$
above these contributions.

We are now in the position of calculating the forward-backward asymmetry for
leptons and the dilepton mass distribution in the SM.  We use
$m_t=175$ GeV for the evaluation of the short-distance coefficients and we take
$|V_{tb}^*V_{ts}|\approx s^2\q_c$. As it can be seen in Fig.~\ref{f_as1}(a),
the SM has a large and negative $A_{FB}$. This is due to the fact that the
largest helicity amplitude is $|H_{+}^{L}|$. The right-handed lepton amplitudes
are negligible, mostly due to cancellations among short-distance coefficients,
whereas $|H_{-}^{L}|$ is suppressed by the combination of form-factors entering
in (\ref{helamp_1}) with $\h_-=1$. The dilepton mass distribution is shown in
Fig.~\ref{f_mll1}(a).

\subsection{Two-Higgs Doublet Model} As a first extension of the minimal SM we
 consider an extension in the Higgs sector. In models with two Higgs doublets,
tree level flavor changing neutral currents are avoided by coupling quarks of
the same charge to the same Higgs doublet. One such model corresponds to the up
quarks getting their masses from one scalar doublet and the down quarks from
the other. The resulting coupling of the charged Higgs to quarks implies a new
contribution to the loop in Fig.~(\ref{f_bsll}) by replacing the $W$ with the
charged Higgs. These contributions have been extensively studied in the
literature \cite{opbasis,grins_bsg}. They effectively shift the values of the
short distance coefficients $C_7$, $C_8$ and $C_9$ at $\m=M_W$. We choose to
study the Model~II in the notation of Ref.~\cite{opbasis}, which corresponds to
the Higgs sector of the minimal supersymmetric standard model. Its short
distance coefficients are given by
\bz
C_7(M_W)=\fr{1}{2}A(x)+G(y)+\fr{1}{6}\left(\fr{v_2}{v_1}\right)^2 A(x)
\label{c7_2hd}
\ez
where
\bz
G(y)=\fr{y}{2(y-1)^2}\left[\fr{5}{6}y-\fr{1}{2}-\fr{(y-2/3)}{(y-1)}\ln y\right]
\label{g_2hd}
\ez
with $x=m_{t}^{2}/M_{W}^{2}$ and $y=m_{t}^{2}/m_{h}^{2}$. The coefficients
$C_8$ and $C_9$ at the electroweak scale are
\by
C_8(M_W)&=&\fr{B(x)}{s^2\q_w}+\fr{4s^2\q_w-1}{s^2\q_w}\left[C(x)-\left(\fr{v_2}{v_1}\right)\fr{x}{2}B(y)\right] \nn \\
& &+D(x)-\fr{4}{9}+\fr{v_2}{v_1}yF(y) \label{c8_2hd}\\
C_9(M_W)&=&-\fr{B(x)}{s^2\q_w}+\fr{1}{s^2\q_w}\left[C(x)-\left(\fr{v_2}{v_1}\right)^2\fr{x}{2}B(y)\right] \label{c9_2hd}
\ey
with
\bz
F(y)=\fr{1}{(y-1)^3}\left[\fr{47}{108}y^2-\fr{79}{108}y+\fr{19}{54}+\fr{(y/3-y^3/6-2/9)}{(y-1)}\ln y \right] \label{f_2hd}
\ez
On the other hand, extending the Higgs sector does not give contributions to
operators outside the SM operator basis (\ref{opbasis}). The coefficients
$C'_7$, $C'_8$ and $C'_9$ corresponding to the operators in (\ref{w_basis})
remain zero. The model is determined by the value of $m_h$, the charged Higgs
mass, and $tan\b=v_2/v_1$, the ratio of vacuum expectation values of the two
Higgs doublets. For instance, for $m_h=200\gev$ and $tan\b=1$ the $A_{FB}$ in
this model is given by the dotted line in Fig.~\ref{f_as1}(a).
The asymmetry is reduced with respect to its SM value. The comparison to the SM
dilepton mass distribution can be seen in Fig.~\ref{f_mll1}(a).

\subsection{TopColor Models} We now turn to a class of models that has the
potential to give not only sizeable shifts in the coefficients $C_7$, $C_8$ and
$C_9$ but also can generate contributions to ``wrong" chirality operators.
Non-zero values of $C'_7$, $C'_8$ and $C'_9$ strongly affect the pattern of
helicities with respect to the SM. This can be clearly appreciated from the
expression (\ref{helamps}) for the transverse helicities, where the terms
containing the ``wrong"-chirality coefficients are multiplied by the
combination of form-factors with the opposite relative sign. This implies, for
instance, that $H_{-}^{L}$ could become comparable to $H_{+}^{L}$, which would
induce a cancellation in $A_{FB}$. Right-handed helicities could also become
important.

TopColor models \cite{chill} are of  interest due to the fact that they give
special status to the third generation through the dynamical generation of the
top quark mass. The dynamics at $\approx 1 $TeV is given by the gauge structure
\bz
SU(3)_1\times U(1)_{Y_1}\times SU(3)_2\times U(1)_{Y_2}\to SU(3)_{\rm
QCD}\times U(1)_{\rm Y} \label{group}
\ez
The $SU(3)_1\times U(1)_{Y_1}$ couples strongly to the third generation whereas
the $SU(3)_2\times U(1)_{Y_2}$ is strongly coupled to the first two.
The $SU(3)_1\times U(1)_{Y_1}$ is assumed to be strong enough to form a chiral
$\langle t\bar{t}\rangle$ condensate, but not $\langle b\bar{b}\rangle$ or
$\langle \t\bar{\t}\rangle$ condensates, giving a large mass to the top quark.
There is a residual $SU(3)'\times U(1)'$ which implies the existence of a
massive color octet $B^{a}_{\m}$ (topgluon) and a singlet $Z'_\m$. The
couplings of the latter to the quarks are given by \cite{chill}
\by
{\cal L}_{Z'}&=&Z'_\m g_1\left\{\tan\q'\left(\fr{1}{3}\bar{\j}_L\g_\m\j_L
+\fr{4}{3}\bar{u}_R\g_\m u_R-\fr{2}{3}\bar{d}_R\g_m d_R\right) \right.\nn \\
& &\left.-\cot\q'\left(\fr{1}{3}\bar{\chi}_L\g_\m\chi_L+\fr{4}{3}\bar{t}_R\g_\m
t_R-\fr{2}{3}\bar{b}_R\g_\m b_R \right)\right\}\label{lzp}
\ey
where $\j=(u,d)$, $\chi=(t,b)$ and  $g_1$ is the  $U(1)_{\rm Y}$ coupling. The
angle $\q'$ is small, which selects the top quark direction for condensation.
The corresponding $Z'_\m$ coupling to first and second generation leptons is
also suppressed by $\tan\q'$, whereas the coupling to the $\t$ lepton is
enhanced by $\cot\q'$. After rotation to the mass eigenstates, (\ref{lzp})
generates four-fermion interactions leading to additional contributions to the
\bqll transitions. For $\ell=e,\m$, the short-distance coefficients are
independent of $\q'$ and are given by
\begin{mathletters}
\label{shift}
\bz
C_8=C_{8}^{SM}+\fr{1}{2}\k \fr{V_{bs}}{V_{tb}V^{*}_{ts}}
 \label{shift_1}
\ez
\bz
C_9=C_{9}^{SM}+\fr{1}{6}\k  \fr{V_{bs}}{V_{tb}V^{*}_{ts}} \label{shift_2}
\ez
\end{mathletters}
and, for the coefficients of (\ref{w_basis})
\begin{mathletters}
\label{new}
\bz
C'_8=-\k \fr{W_{bs}}{V_{tb}V^{*}_{ts}}\label{new_1}
\ez
\bz
C'_9= -\fr{1}{3}\k\fr{W_{bs}}{V_{tb}V^{*}_{ts}}\label{new_2}
\ez
\end{mathletters}
where we have defined
\bz
\k=\fr{8\p^2 v^2}{M_{Z'}^{2}}\left(\fr{M_Z}{M_W}\right)^2 \label{kappa}
\ez
and $V_{bs}=D^{*L}_{bs}D^{L}_{bb}$ and $W_{bs}=D^{*R}_{bs}D^{R}_{bb}$ are the
residual non-diagonal terms in the $Z'$ couplings after the diagonalization of
the mass matrix for down-type quarks according to  $D^{L\dagger}M_DD^R$. The
FCNC couplings $V_{bs}$ and $W_{bs}$ are only fixed  by a specific realization
of the model. In order to make a prediction, we take them to be equal and
approximately half the SM charged current equivalent $V_{ts}$ (The squared root
anzatz in \cite{chill}).
\bz
|V_{bs}|\approx |W_{bs}|\approx \pm \fr{1}{2} |V_{ts}| \label{sqr_anz}
\ez
The results for the asymmetry are plotted in Fig.~\ref{f_as1}(b) for the minus
sign in (\ref{sqr_anz}) and for both $M_{Z'}=500 \gev$ and $M_{Z'}=1 {\rm TeV}$
. In both cases, but in particular in one of them, it can be seen that the
asymmetry is very different than the one obtained in the SM. The dilepton mass
distributions are shown in Fig.~\ref{f_mll1}(b). The effect of the $Z'$ is
striking both in the asymmetry and the dilepton mass distribution, even when
looking at such a restricted region of phase space. The asymmetry becomes a
more sensitive test for a much heavier $Z'$. For instance for $M_{Z'}=1$ TeV
the branching ratio is not much larger than in the SM, whereas the asymmetry is
still significantly different.
In Table~I we show the branching ratios and asymmetries corresponding to $\mll
>3.8 \gev$ in the cases considered above.

\noindent  We finally point out that TopColor models as the ones considered
above, that produce observable effects with $\ell=e$  and  $\m$ are likely to
give very large effects for $\ell=\t$. This is due to the fact that the
four-fermion interactions induced by (\ref{lzp}) together with the similar
Lagrangian for $\t$ leptons, are proportional to $\cot^2 \q'$. This implies
large contributions to $b\to s\t^+\t^-$ processes like $B\to K^{(*)}\t^+\t^-$
and $B_s\to\t^+\t^-$ \cite{gb}.

The results for \bk can be seen in Fig.~\ref{f_mll2}, where the dilepton mass
distributions are plotted for the various cases considered above
and in the region allowed by the use of the relations (\ref{scal_p}). It can be
seen that this mode is not such a powerful test of the SM. However its
observation will be an important complement to the information extracted from
\bks. The results for this decay are summarized in Table~II.

\section{Conclusions}
\label{conc}
\begin{figure}
\vspace{7cm}
\includegraphics{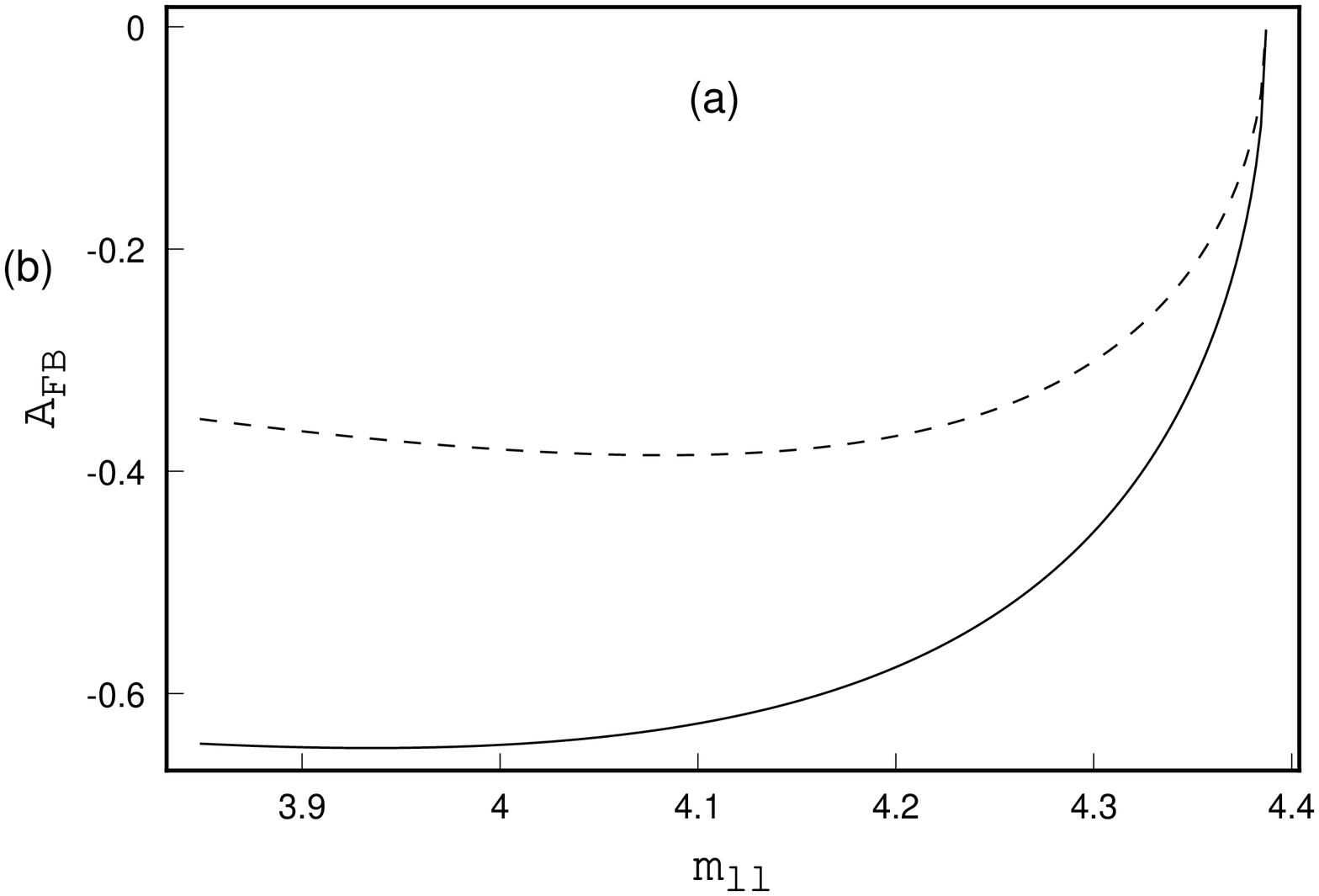}
\includegraphics{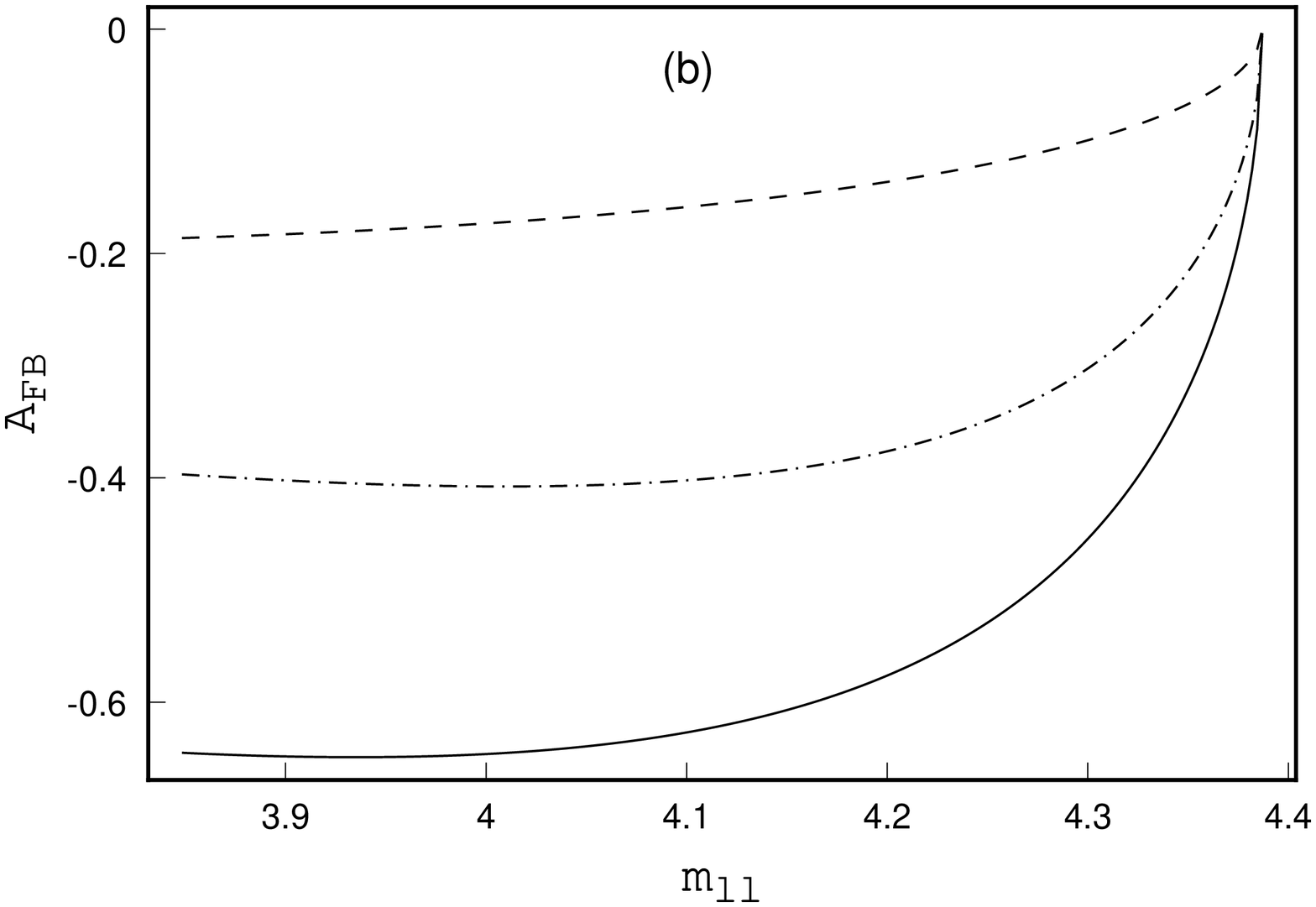}
\vspace{0.0cm}\hspace{3cm}\hspace{8.5cm}\vspace{1cm}
\vspace{7cm}
\caption[]{ Forward-Backward asymmetry for leptons as a function of the
dilepton mass in \bks. The solid line is the SM prediction. In (a), the dashed
line is the Two-Higgs doublet model. In (b), the dashed line is the Top-Color
prediction for $m_{Z'}=500$ GeV, whereas the dotted-dashed line corresponds to
$m_{Z'}=1$ TeV. Both cases correspond to the square root anzatz of
eqn.~(\ref{sqr_anz}) with a negative relative sign.  }
\label{f_as1}
\end{figure}

\begin{figure}
\vspace{7cm}
\includegraphics{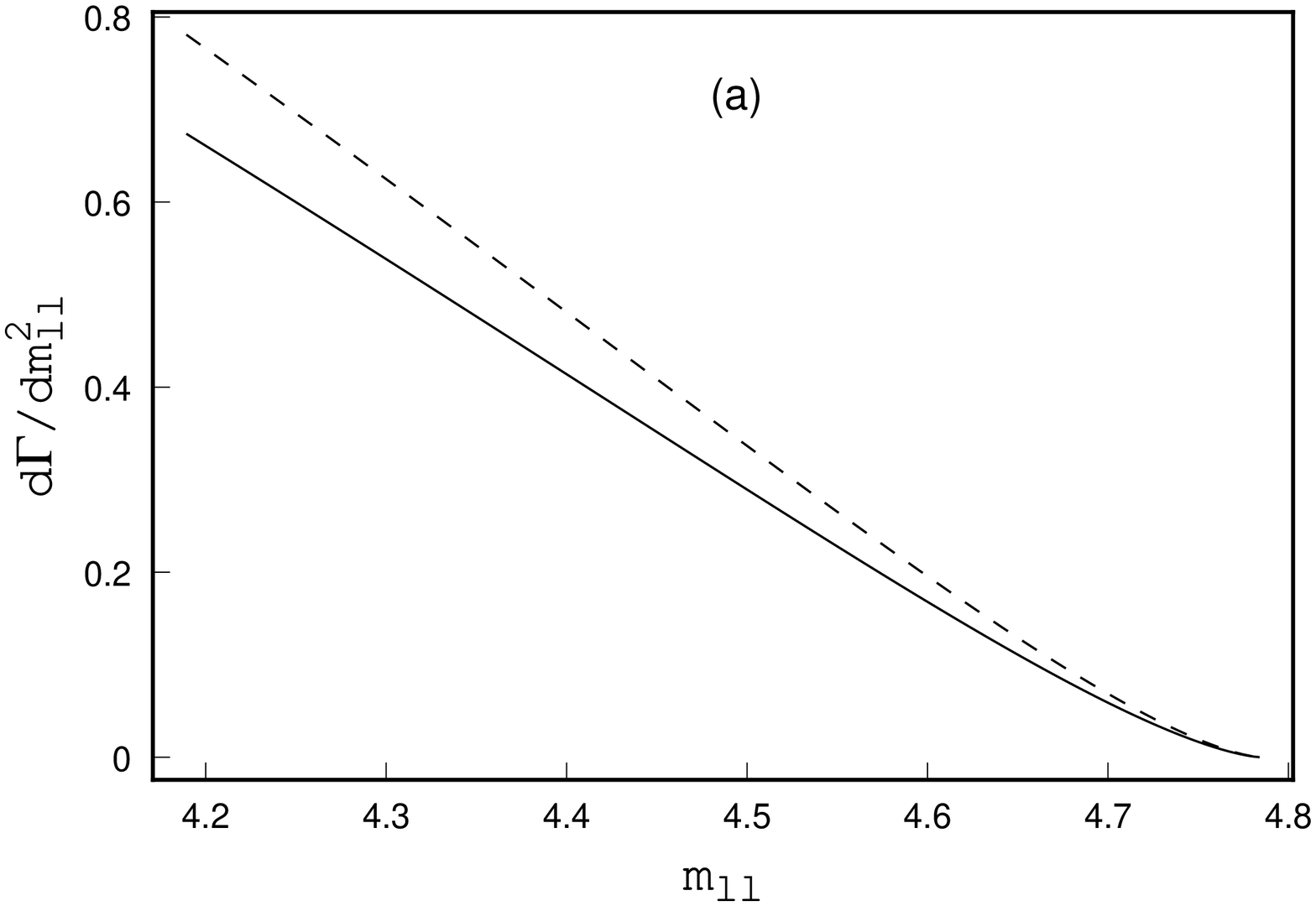}
\includegraphics{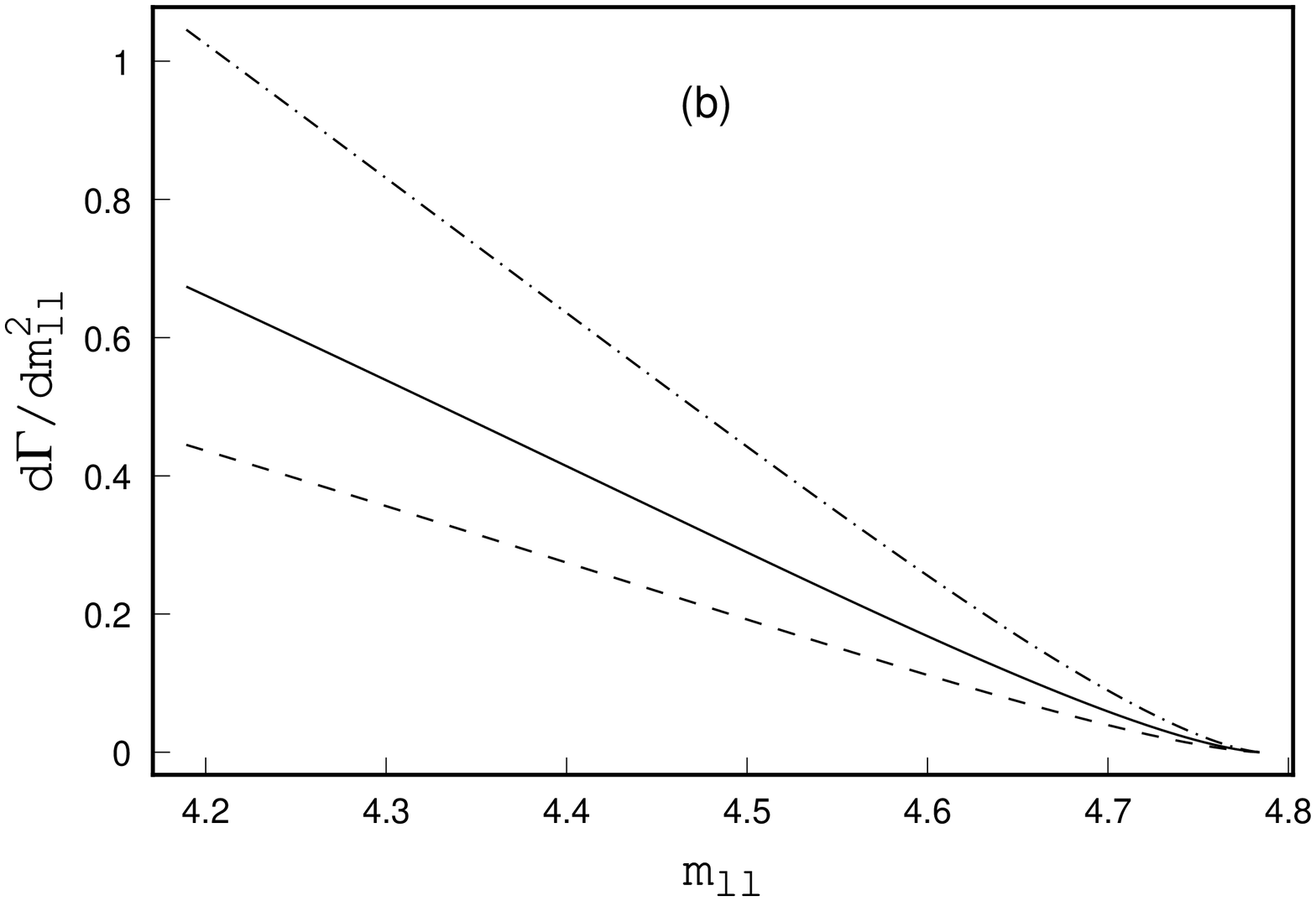}
\vspace{0.0cm}\hspace{3cm}\hspace{8.5cm}\vspace{1cm}
\vspace{7cm}
\caption[]{Dilepton mass distributions in \bk. The caption is the same as in
Fig.~\ref{f_as1}. }
\label{f_mll2}
\end{figure}

As we have seen in previous sections, the decay \bks is a powerful test of the
SM. This is true even  when the present experimental situation restricts the
reliable predictions of hadronic matrix elements to the region with
$\mll^2>m_{\Psi'}^{2}$. The angular information is likely to be an extremely
useful tool given its sensitivity to changes in the short distance coefficients
in (\ref{heff}). The forward-backward asymmetry $A_{FB}$ is very large and
negative in the SM, as it can be seen in Fig.~\ref{f_as1}(a) and in Table~I.
This is caused by an accidental partial cancellation in the short-distance
factor
that makes the helicity $H_{+}^{L}$ to be  significantly larger than all the
others, causing a large negative value in (\ref{afb_hel}).

\begin{figure}
\vspace{7cm}
\includegraphics{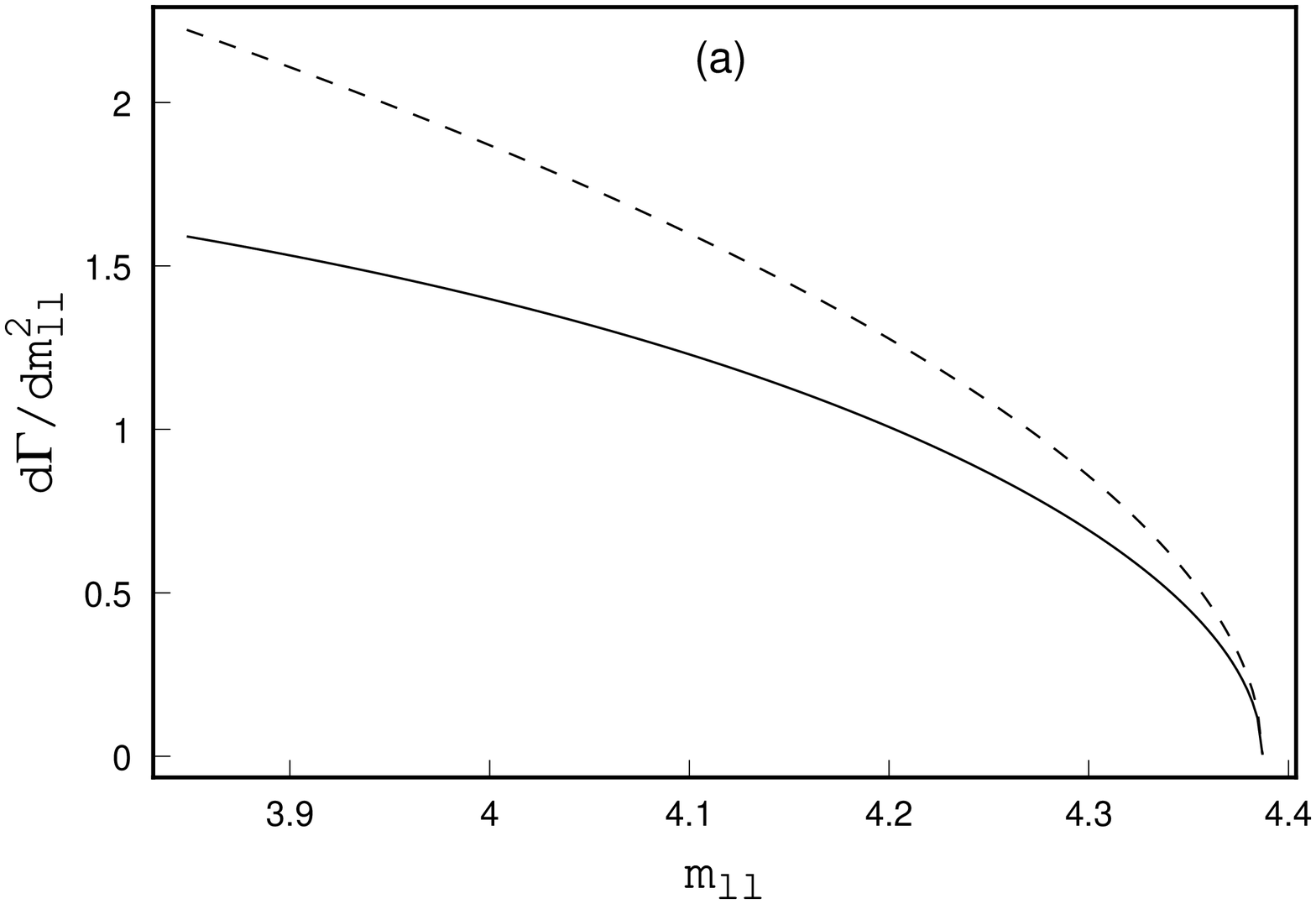}
\includegraphics{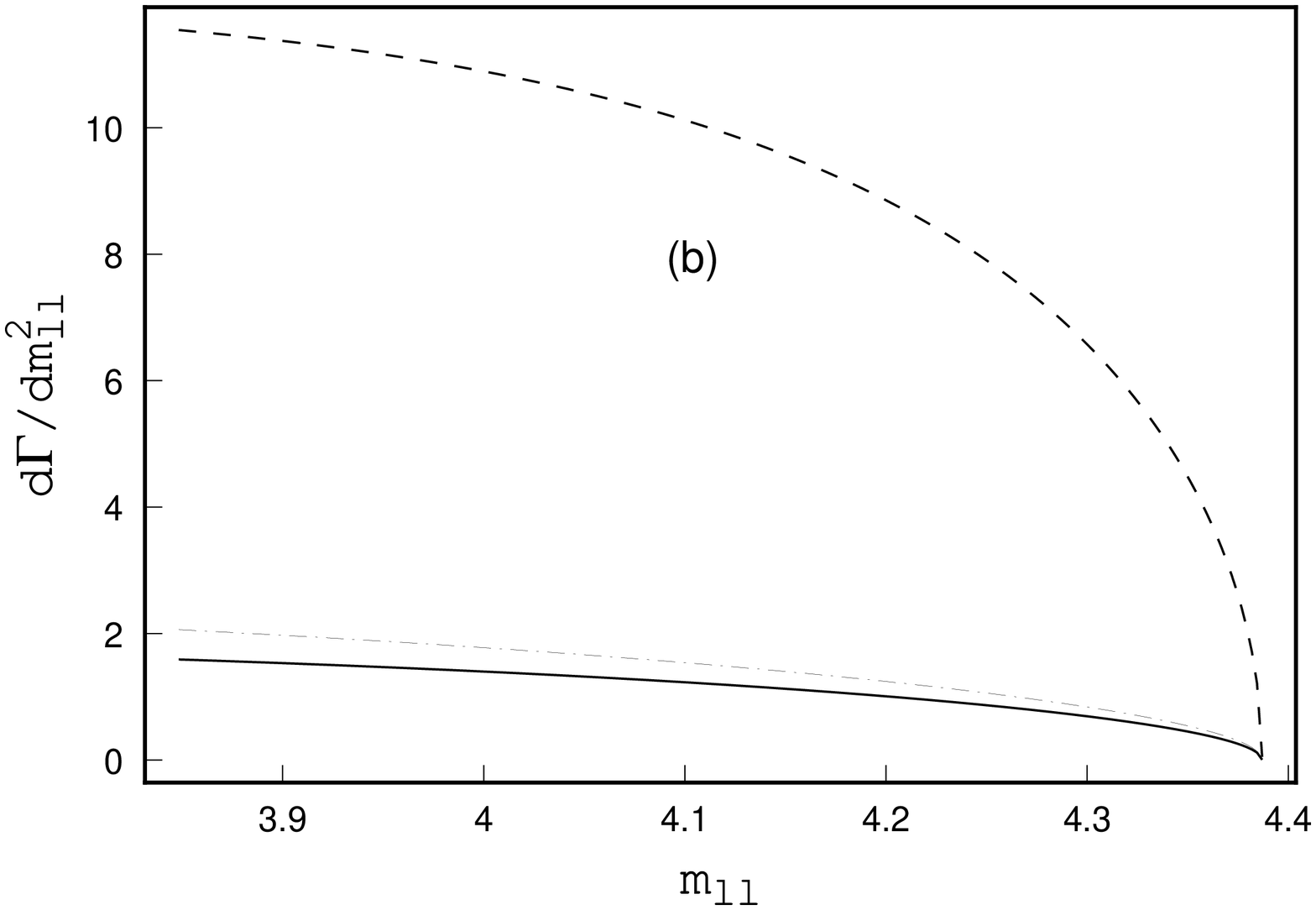}
\vspace{0.0cm}\hspace{3cm}\hspace{8.5cm}\vspace{1cm}
\vspace{0.0cm}
\caption[]{Dilepton mass distributions for \bks. The caption is the same as in
Fig.~\ref{f_as1}. }
\label{f_mll1}
\end{figure}

On the other hand, the Two-Higgs doublet scenario we discussed in
Sect.~\ref{predictions}, with $\tan\b=1$ and  $m_h=200$ GeV, gives a
considerably smaller asymmetry even when the dilepton mass distribution is not
very different from that of the SM (Fig.~\ref{f_mll1}(a)).

 We have also considered the effects of TopColor models \cite{chill}. The \bqll
decays are the first low energy data that could be strongly affected by the
phenomenology of these models, which are also expected to produce important
effects on top and bottom production \cite{hill_parke} as well as on $\G(Z\to
b\bar{b})$ \cite{hill_zhang}. We have seen in Sect.~\ref{predictions} that a
$500$ GeV color-singlet boson strongly coupled to the third generation gives
large deviations from the SM in both the asymmetry and the dilepton mass
distribution, as long as the neutral mixing induced is chosen to be of the same
order as the corresponding CKM matrix element as in (\ref{sqr_anz}). The
asymmetry (Fig.~\ref{f_as1}(b), dashed line) is largely reduced with respect to
the SM value. This is due not only to shifts in the coefficients $C_8$ and
$C_9$ which upset the cancellation taking place in the SM, but also due to
non-zero values of $C'_8$ and $C'_9$, the coefficients of the ``wrong"
chirality operators $O'_8$ and $O'_9$ in (\ref{w_basis}). The latter implies
important contributions to helicity amplitudes that were largely suppressed in
the SM. The sensitivity of the angular information to non-zero values of these
coefficients is illustrated in the case $m_{Z'}=1$ TeV. There, the dilepton
mass distribution (Fig.~\ref{f_mll1}(b), dotted-dash line) is not very
different from the SM prediction. However, the asymmetry is still substantially
smaller, as can be appreciated from Fig.~\ref{f_as1}(b).

Although is true that reliable predictions for the hadronic matrix elements are
presently limited to large dilepton masses, this might not be such an important
limitation regarding the quality of the angular information. The availability
of $B\to\rho\ell\n$ data will allow SM tests for all dilepton masses in \bks.
However the distinctive features of the angular information are likely to
disappear or not be so important for smaller $\mll^2$. For instance, the SM
asymmetry will not be so large for smaller $\mll^2$ due to the absence of a
near cancellation in some of the short distance factors in (\ref{helamps}).
This suggests that concentrating on the region of large $\mll^2$ is not just
convenient because of the lack of information from $B\to\rho\ell\n$, but also
because the angular information is most discriminating there. The pseudoscalar
mode \bk is not as rich in information as \bks is, but it can be a powerful
complement to it. For instance, the Two-Higgs doublet model considered here
gives slightly larger rates in both modes, whereas the TopColor model that
gives the largest rate in \bks produces  a rate in \bk that is {\em smaller}
than in the SM, as can be seen in Table~II.

In the near future experiments in both hadron and $e^+e^-$ colliders will have
sensitivity to branching fractions as the ones in Tables~II and II. The current
upper limit corresponds to a partial branching fraction in the region we have
considered, of approximately $(1-2)\times 10^{-5}$  \cite{cleo_2,cdf} for the
\bks mode. Therefore, experimental information is about to start constraining
new physics models and soon will be at the level of the SM predictions.
The reconstruction of the dilepton asymmetries will then be an important probe
into the physics of the electroweak symmetry breaking scale.

\acknowledgments The author thanks Marcia Begalli, Gerhard Buchalla, Fritz De
Jongh and Chris Hill for useful suggestions and comments. This work was
supported by the U.S. Department of Energy.

\ls{1.0}
\begin{table}
\begin{tabular}{ccc}
&$a_{FB}^{\em partial}$&$B.R.^{\em partial}(\times 10^{-6})$\\ \tableline
SM&$-0.57$&$0.53$\\
2HD&$-0.35$&$0.64$\\
TopC ($M_{Z'}=500\gev$)&$-0.138/-0.250$&$4.0/5.0$\\
TopC ($M_{Z'}=1$ TeV)&$-0.36/-0.46$&$0.6/0.9$
\end{tabular}
\vskip 0.75truecm
\caption{Integrated asymmetry and branching fraction in the region $\mll>3.8
\gev$ in \bks. For TopColor we take into account both possible relative signs
$(-/+)$ between the FCNC couplings and the corresponding CKM matrix elements.}
\end{table}

\vskip2truecm

\begin{table}
\begin{tabular}{cc}
&$B.R.^{\em partial}(\times 10^{-7})$\\ \tableline
SM&$1.6$\\
2HD&$1.9$\\
TopC ($M_{Z'}=500\gev$)&$1.1/2.5$\\
TopC ($M_{Z'}=1$ TeV)&$0.6/1.0$
\end{tabular}
\vskip 0.75truecm
\caption{Integrated  branching fraction in the region $\mll>4.2 \gev$ in \bk.
For TopColor we take into account both possible relative signs $(-/+)$ between
the FCNC couplings and the corresponding CKM matrix elements.}
\end{table}

\newpage

\eject

\newpage

\ls{1.0}

\end{document}